\documentclass{PoS}
\def\Pom{{\bf I\!\!P}}
 
\title{Central exclusive photoproduction at the LHC} 
 
\ShortTitle{Central exclusive photoproduction at the LHC} 
 
\author{\speaker{Wolfgang SCH\"AFER}%
        \thanks{Partially supported by the Polish Ministry of Science and Higher Education under contract 1916/B/H03/2008/34.}\\ 
        IFJ PAN, Krak\'ow\\ 
        E-mail: \email{wolfgang.schafer@ifj.edu.pl}} 
 
\author{Anna CISEK\\ 
        IFJ PAN, Krak\'ow\\ 
        E-mail: \email{anna.cisek@ifj.edu.pl}} 

\author{Antoni SZCZUREK\\ 
        IFJ PAN, Krak\'ow and Rzesz\'ow University\\ 
        E-mail: \email{antoni.szczurek@ifj.edu.pl}} 
 
\abstract{
  Protons and antiprotons at collider energies are a source of high energy 
Weizs\"acker--Williams photons. This opens up a possibility to study at the LHC exclusive photoproduction of heavy vector mesons at energies much larger 
than possible at the HERA accelerator. 
We present selected results of our detailed studies of various distributions 
for the production of heavy quarkonia 
(e.g. rapidity, transverse momenta, azimuthal angles). 
We give predictions for LHC energies. 
Our calculations are based on modelling the photoproduction amplitudes 
in a $k_\perp$-factorisation approach which is checked against HERA data. 
We also discuss the exclusive photoproduction of $Z^0$ bosons.
} 
 
\FullConference{European Physical Society Europhysics Conference on 
High Energy Physics\\ 
           July 16-22, 2009\\ 
           Krakow, Poland} 
 
\begin{document} 
 
Central exclusive production processes have recently attracted 
much attention for offering a rich physics program at the
modern colliders, from a potentially clean way to discover/investigate 
a light Higgs to (exotic) hadron spectroscopy to 
investigations of the QCD Pomeron (for recent reviews with many references, 
see \cite{Reviews}).

Here we show a selection of results obtained in \cite{SS07,RSS08,CSS09}.
We concentrate on a particular production mechanism: photoproduction.
In the language of $t$-channel exchanges, we describe it in terms of the
photon--Pomeron fusion. We want to have large rapidity gaps, so the exchange
mechanism must have an effective spin $\geq 1$, and must not 
carry neither charge nor color charge. 
Those requirements are fulfilled by both photon and Pomeron.
The $C$-parity of the centrally produced system must be the product
$C = C_\gamma \cdot C_\Pom = -1$. Hence we will concentrate on the production
of (heavy) vector mesons $J/\psi,\Upsilon$ and their radial excitations,
and we will also show predictions for the electroweak $Z^0$ gauge boson.

Ideally one would measure the final state completely, including the 
protons. An interesting feature of the photon exchange
mechanism is that the interference between the photon-Pomeron fusion,
and the exchange-diagram Pomeron-photon fusion, induces an azimuthal
correlation between the outgoing protons \cite{SS07}.
In practice, if protons are not measured, one has to also include 
low--mass diffractive excitation in the final state. 
A review of the experimental situation at the LHC has been given
at this conference \cite{experiment}. 
Recently, the experimental proof of principle has been given by CDF at
the Tevatron, where exclusive charmonium production has been
measured for the first time in hadronic collisions at high energy
\cite{CDF_Jpsi}.

\begin{figure}
\includegraphics[width= \textwidth]{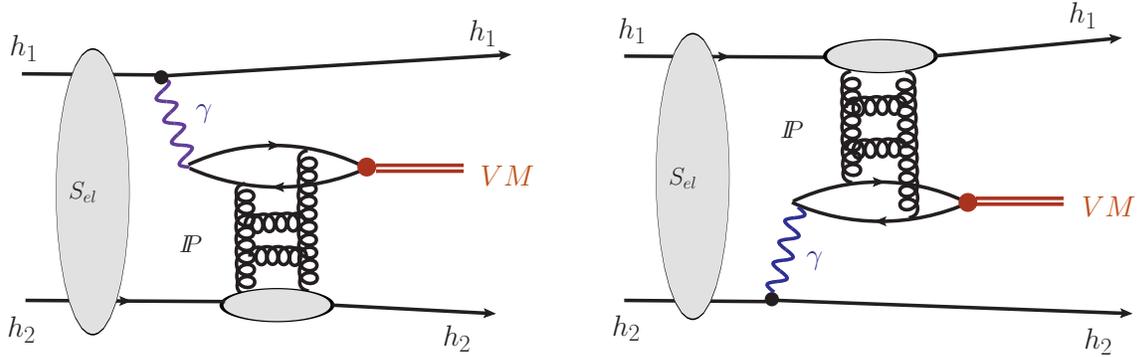}
\caption{The photon-Pomeron fusion mechanism. The Pomeron exchange
is modelled by a pQCD gluon ladder. Also shown is the absorptive 
correction, including elastic rescattering. 
}
\label{fig1}
\end{figure}

\begin{figure}
\includegraphics[width=.5\textwidth]{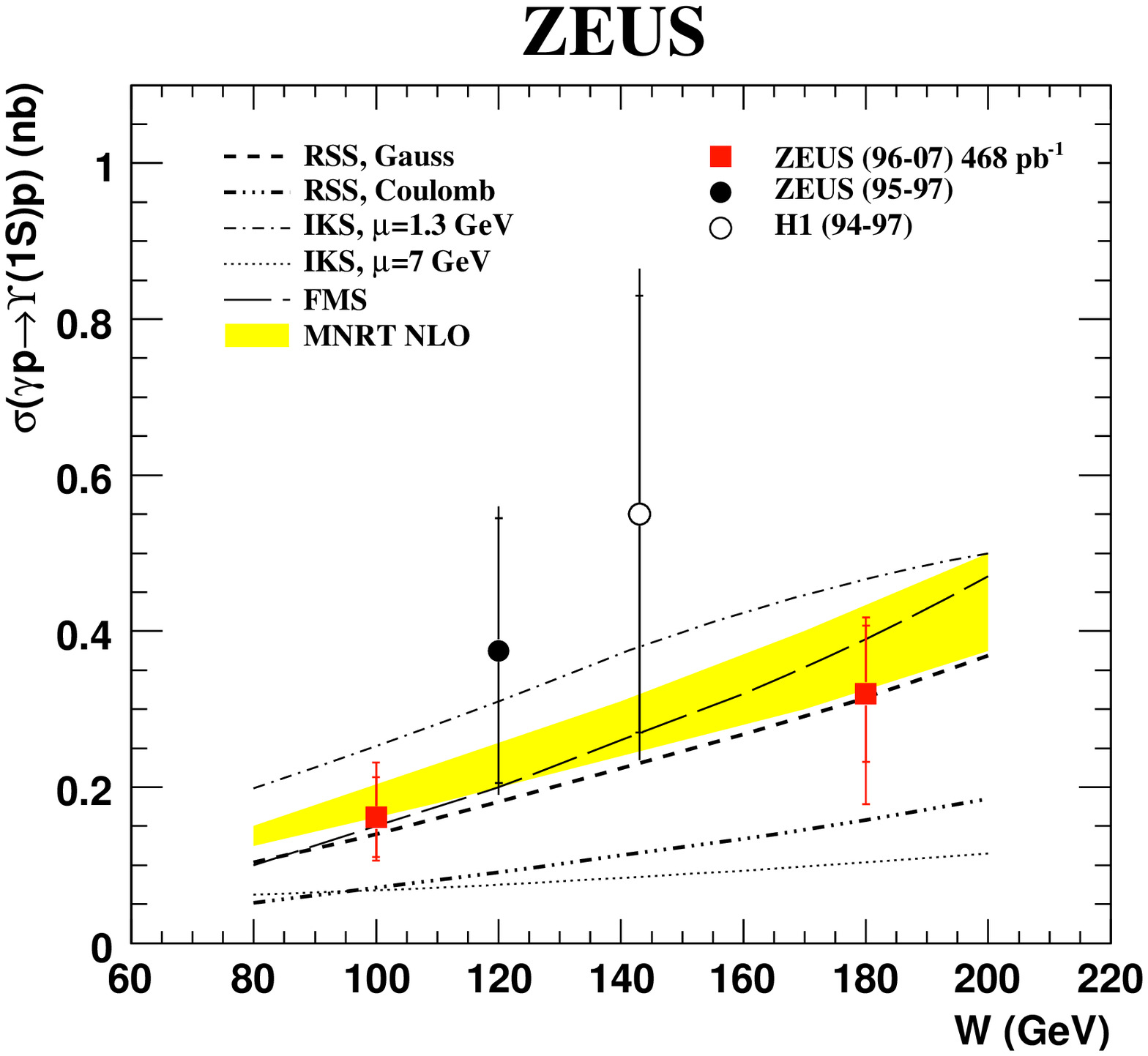}
\includegraphics[width=.6\textwidth]{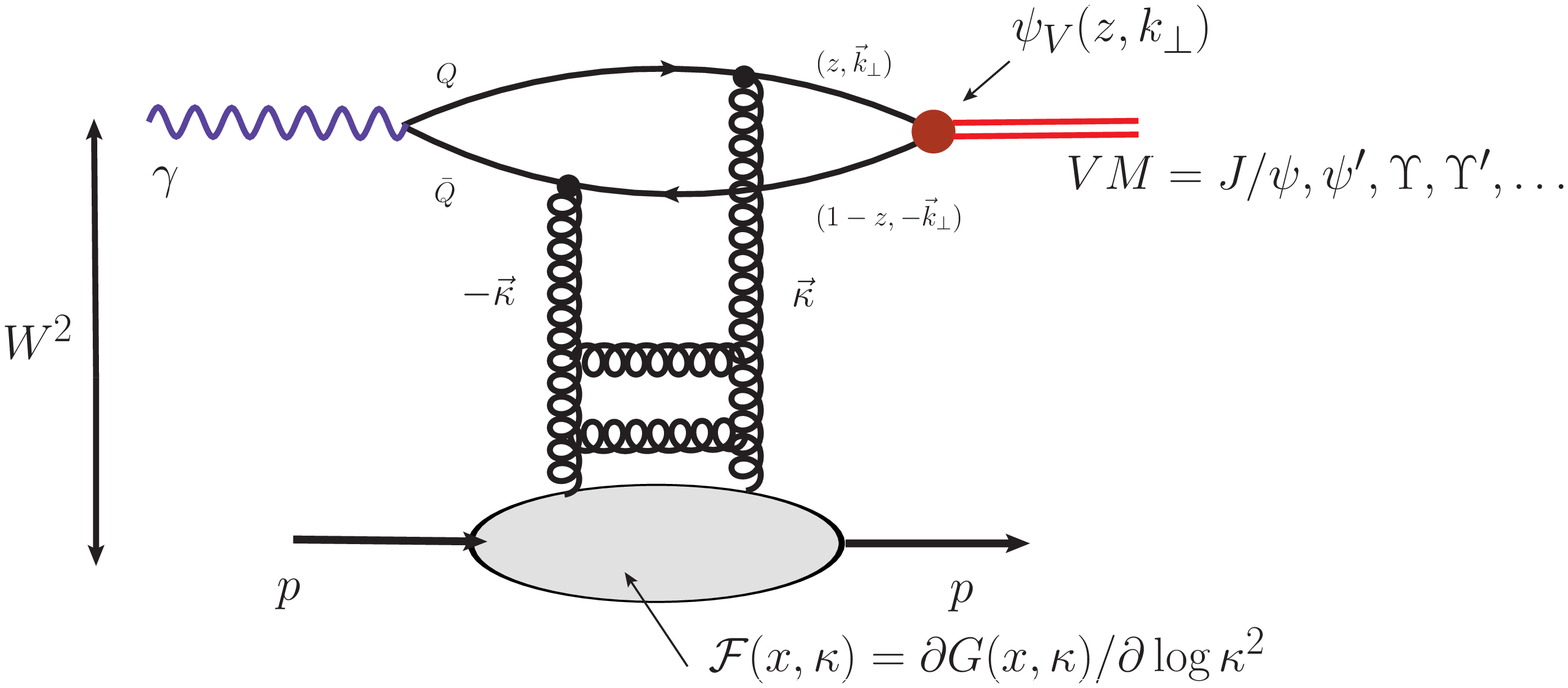}
\caption{Left: the total cross section for $\gamma p \to \Upsilon p$
taken from \protect{\cite{ZEUS}}. Our results are labelled RSS.
Right: the pQCD model for the production amplitude. The input unintegrated
gluon distribution is constrained by HERA data, the vector meson wavefunction
has to reproduce the leptonic decay width.}
\label{fig2}
\end{figure}

With respect to other possible hadronic exchange mechanisms, such as
the conjectured Odderon \cite{Odderon}, the photon exchange
stands out through the very sharp forward peaking induced by the photon
pole. Then the relevant photon virtualities are small enough 
to limit ourselves to quasireal, Weizs\"acker--Williams photons.
We are dealing with predominantly peripheral collisions, and 
can expect the effect of absorptive corrections to be only on the 
$\sim 10 \div 20 \%$--level.  
The diagrams fig \ref{fig1} can then be calculated essentially 
from the amplitude of vector-meson photoproduction on a nucleon. 
The latter is strongly constrained by data obtained at HERA.
In fig. \ref{fig2} our results for the total photoproduction 
cross section of $\Upsilon$ are shown against recent data 
from the ZEUS collaboration (from \cite{ZEUS}). For a specific
choice of the vector meson wavefunction the agreement is quite good. 
The calculation is based on a pQCD modelling of Pomeron exchange 
along the lines of \cite{INS}, for more details, see \cite{RSS08}.

\begin{figure}
\includegraphics[width=.5\textwidth]{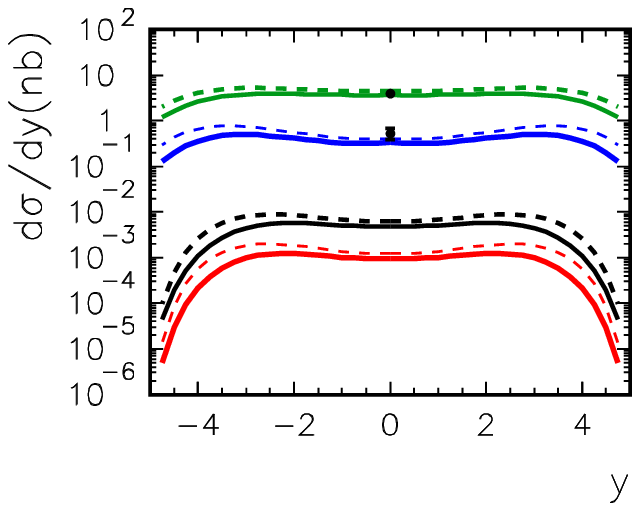}
\includegraphics[width=.5\textwidth]{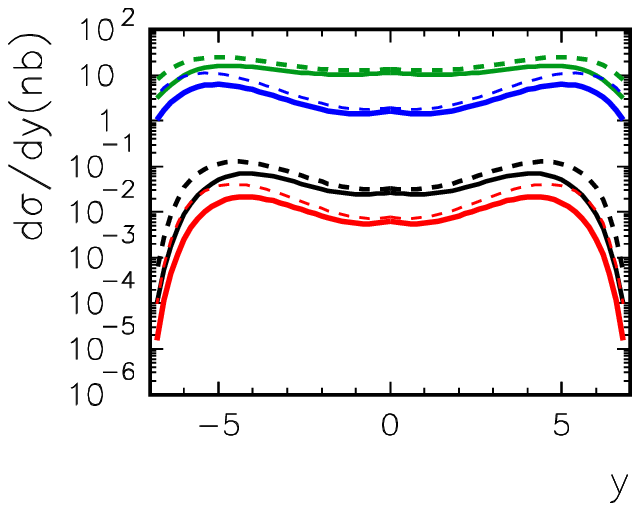}
\caption{Rapidity spectrum of exclusive vector mesons for Tevatron, 
$\sqrt{s} = 1960$ GeV, (left) and LHC, $\sqrt{s} = 14$ TeV, (right) energies. 
Green: $J/\psi$, Blue:$\psi'(2S)$, Black:$\Upsilon$, Red:$\Upsilon(2S)$. 
Notice that at LHC energies for $\Upsilon$ at $y \sim 0$ we probe the glue
at $x \sim 10^{-3} \div 10^{-4}$, while at $y \sim 5$ effectively 
$x \sim 10^{-5} \div 10^{-6}$. The data points are from 
{\protect \cite{CDF_Jpsi}}. Dashed lines are without, solid lines 
with absorption.}
\label{fig3}
\end{figure}

In fig. \ref{fig3} we show rapidity distributions for various 
vector mesons in proton-proton (or proton antiproton)
collisions. We also show the recent CDF data \cite{CDF_Jpsi}, the
agreement being quite reasonable. It is interesting to observe
that the effect of absorptive corrections varies over phase space.
Born--level results are shown by the dashed lines, and the full
(including absorptive corrections) results by the solid lines.
Absorption is a stronger effect at larger rapidities. This is related
to the fact that there a larger longitudinal momentum transfer
is involved, and hence the collision is less pripheral, and 
therefore absorption is stronger. It is important to keep
this in mind if one wants to draw conclusions on the energy dependence
of the $\gamma p \to V p$ photoproduction amplitude from the
shape of the rapidity distribution.
As far as the production of radially excited vector mesons 
is concerned, they will be produced in roughly the 
same ratio as in the photoproduction on the nucleon target.

\begin{figure}
\includegraphics[width=.5\textwidth]{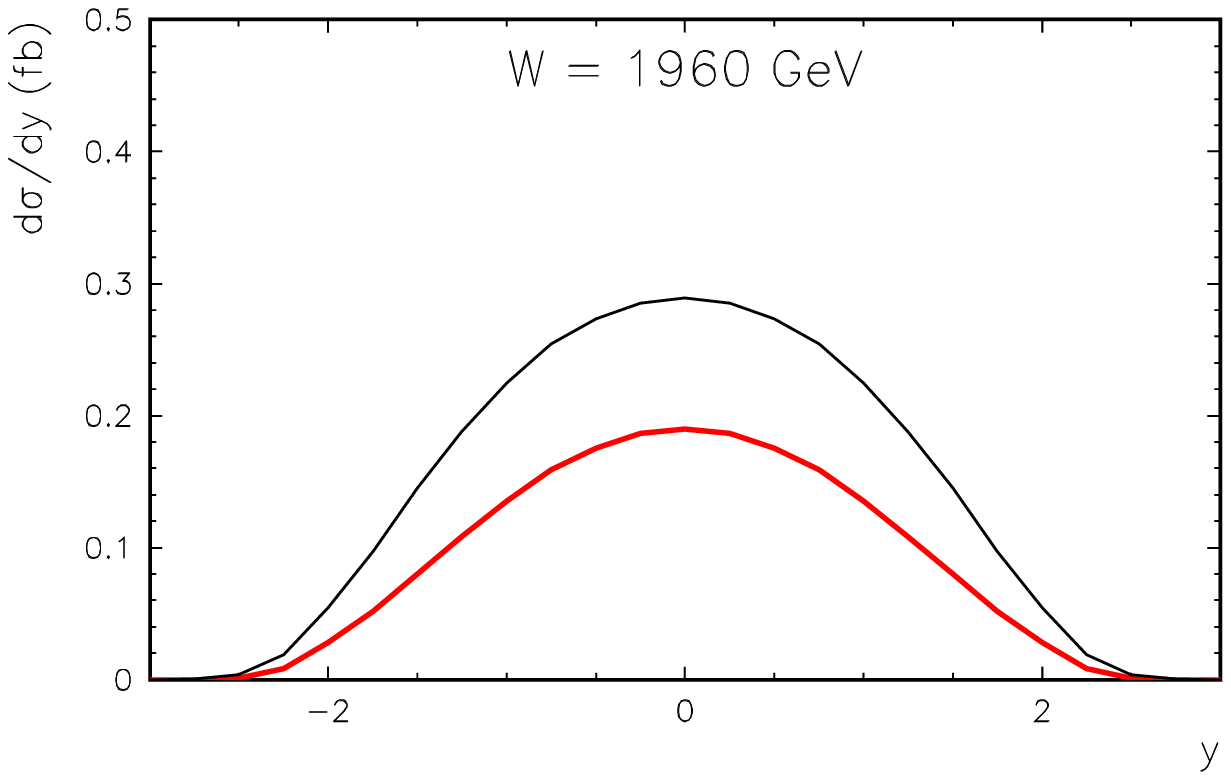}
\includegraphics[width=.5\textwidth]{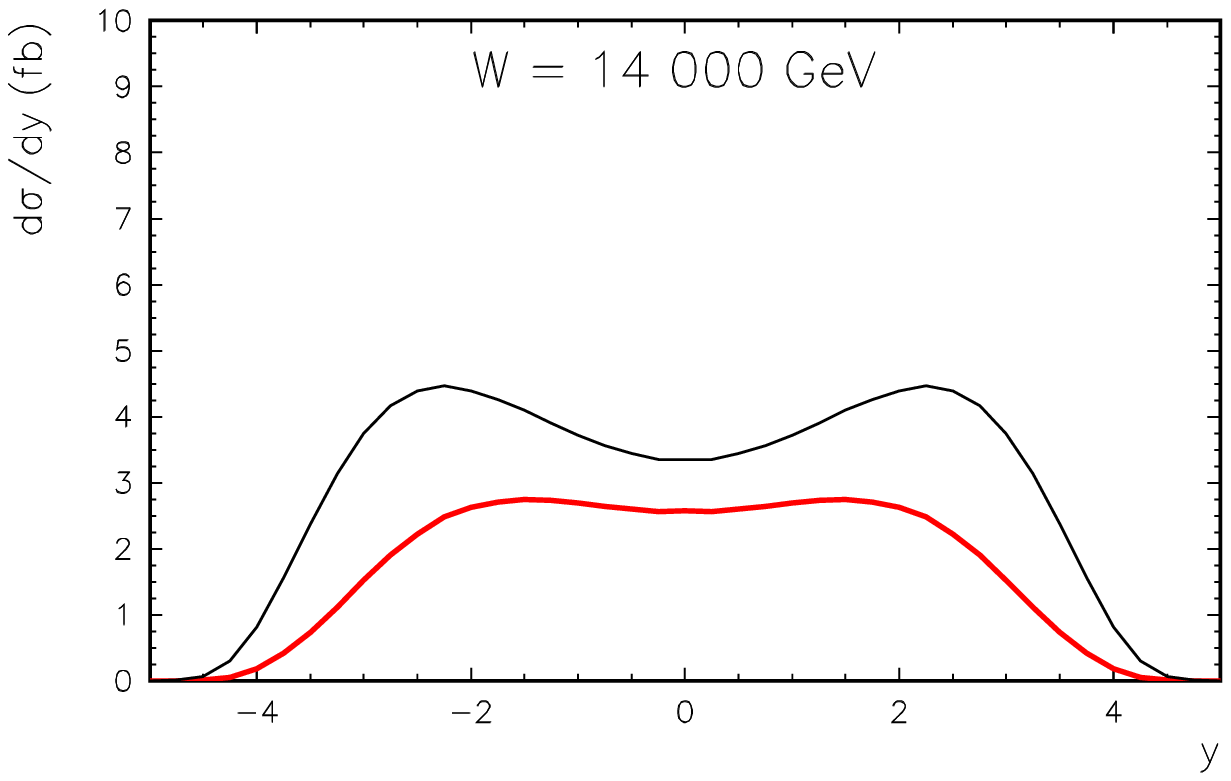}
\caption{Rapidity spectrum of exclusive $Z^0$'s 
at Tevatron (left) and LHC (right) 
energies. The thin black lines are without, the thick 
red lines with absorption.}
\label{fig4}
\end{figure}

Let us now turn to the $Z^0$ production \cite{CSS09}. 
Here there are some subtleties
in the calculation of the photoproduction amplitude, due to
the fact the amplitude has a cut stemming from the $Z^0 \to q \bar q$
transition, differently from the vector meson case. 
The photoproduction cross section at HERA energies is exceedingly small,
indeed this process has never been observed.
In fig. \ref{fig4} we show rapidity distributions of $Z^0$ in $p \bar p$ and
$p p$ collisions. The thin solid line shows the Born level results, while 
for the red curves absorption has been included. 
Absorption effects are stronger than for vector meson production:
the heavier $Z^0$ boson is produced in substantially less peripheral
collisions. The cross section is however disppointingly small.
A recent search by CDF \cite{CDF_Z} came up with an upper limit for the
total cross section of $\sim 1 \, \mathrm{pb}$, roughly a factor thousand
bigger than our result.

In summary, cross sections for exclusive vector mesons are of 
measurable size. The fair agreement of our predictions
with Tevatron data suggests that absorption is under reasonable control. 
A possible measurement at the LHC promises access to the gluon distribution
at very small $x$, beyond the HERA domain.

\end{document}